\documentclass[journal]{IEEEtran}

\usepackage[utf8]{inputenc}
\usepackage{amsmath, amssymb, mathrsfs, eucal, graphicx}
\usepackage[noadjust]{cite}
\usepackage[uline]{hhtensor}

\newtheorem{theorem}{Theorem}[section]
\newtheorem{condition}[theorem]{Condition}
\newtheorem{proposition}[theorem]{Proposition}

\newtheorem{remark}[theorem]{Remark}
\newtheorem{lemma}[theorem]{lemma}

\renewcommand{\IEEEQED}{\IEEEQEDopen}

\newenvironment{enumrm}{\begin{enumerate} \renewcommand{\theenumi}{\roman{enumi}}} {\end{enumerate}}

\begin{document}

\title{Sequential detection of Markov targets with trajectory estimation}
\author{Emanuele~Grossi,~\IEEEmembership{Member,~IEEE},~and~Marco~Lops,~\IEEEmembership{Senior Member,~IEEE}%
\thanks{The authors are with the DAEIMI, Universit\`a degli Studi di Cassino, via G. Di Biasio 43, Cassino (FR), 03043 Italy. E-mail: \{e.grossi, lops\}@unicas.it.}
\thanks{Manuscript received January 26, 2007; revised December 4, 2007.}}
\markboth{Transaction on Information Theory,~Vol.~X, No.~X,~Month~Year}{Shell \MakeLowercase{\textit{et al.}}: Bare Demo of IEEEtran.cls for Journals}

\maketitle

\begin{abstract}
The problem of detection and possible estimation of a signal generated by a dynamic system when a variable number of noisy measurements can be taken is here considered. Assuming a Markov evolution of the system (in particular, the pair signal-observation forms a hidden Markov model), a sequential procedure is proposed, wherein the detection part is a sequential probability ratio test (SPRT) and the estimation part relies upon a maximum-a-posteriori (MAP) criterion, gated by the detection stage (the parameter to be estimated is the trajectory of the state evolution of the system itself). A thorough analysis of the asymptotic behaviour of the test in this new scenario is given, and sufficient conditions for its asymptotic optimality are stated, i.e. for almost sure minimization of the stopping time and for (first-order) minimization of any moment of its distribution. An application to radar surveillance problems is also examined.
\end{abstract}

\section{Introduction}
\IEEEPARstart{M}{any} statistical decision problems in engineering applications require to perform state estimation of a dynamic system under uncertainty as to signal presence \cite{Mid1968, Fre1972, Bay1995}. This includes fault detection and diagnosis in a dynamical system control \cite{Wil1976, Mas1989}, target detection and tracking \cite{Im1999}, image and speech segmentation \cite{Mer1991}, speaker identification and source separation, blind deconvolution of communication channels. Application of sequential decision rules to the above scenario arouses much interest since it promises a considerable gain in sensitivity, measured by the reduction in the average sample number (ASN), with respect to fixed sample size (FSS) procedures. These advantages are particularly attractive in remote radar surveillance, where the signal amplitude is weak compared to the background noise and stringent detection specifications can be met only by processing multiple frames as in \cite{Bar1985, Arn1993, Ton1996}). In this case, FSS techniques usually result to be inefficient while sequential procedures are known to increase the sensitivity of power-limited systems or, alternatively, to reduce the ASN.

The adoption of sequential procedures, however, poses some difficulties: since the instant when the procedure stops sampling is not determined in advance (it is a random stopping time, indeed) the set of trajectories of the dynamic system to be considered (i.e. the parameter space) has an infinite cardinality. On the other hand, sequential testing rules have been already extended to the case of composite hypotheses. In \cite{Mar1962} an SPRT is adopted in a radar framework assuming a prior on the parameter space, in turn consisting of a finite number of elements (the radar resolution cells). Sub-optimal sequential classification procedures (also called multi-hypotheses tests) were also proposed during the past years, such as \cite{Arm1950, Kie1963, Lor1972, Lor1977, Hel1962, Ree1963, Bau1994, Mar2003} for the case of independent and identically distributed (i.i.d.) observations and \cite{Lai1981, Tar1999, Dra1999, Dra2000, Tar2003} for the more general setting of non i.i.d. observations. However, all of these studies were restricted to a finite cardinality of the parameter space, an overly restrictive condition, which corresponds to requiring that the dynamic system may only lie in a determined state, with no transition allowed. Few works in the past have studied sequential problems for hidden Markov models (HMMs), which are known to admit a dynamical system representation in the sense of control theory \cite{Eph2002}. In \cite{Fuh2003} the performances of SPRTs for model estimation in parametrized HMMs and the cumulative sum (CUSUM) procedure for change point detection in HMMs are studied, while \cite{Che2000} addresses the quickest detection of transient signals represented as HMMs using a CUSUM-like procedure, with possible applications to the radar framework.

This paper addresses the problem of sequential detection and trajectory estimation of the state evolution of a dynamical system observed through noisy measurements. In the above framework, its contributions can be summarized as follows.
\begin{itemize}
\item[-] At the design stage, a sequential procedure is defined with no restriction as to the parameter space cardinality. The detection part of the procedure realizes an SPRT while, in order to estimate the system state trajectory, a gated estimator is defined, in the sense that estimation is enabled by the result of the detection operation.
\item[-] It is known that Wald's SPRT for testing simple hypotheses based on i.i.d. observations has a number of remarkable properties \cite{Bur1963, Fer1967}, the most appealing being the fact that it simultaneously minimizes the expected sample size under both hypotheses. These properties, however, fail to hold when the observations are not i.i.d., as it happens when they are generated by a dynamic system. In this paper, a deep asymptotic analysis for the detection part is given and sufficient conditions under which these properties hold are stated, consistent with previous results in \cite{Lai1981, Tar1999, Dra2000}. In particular, it is shown that under a set of rather mild conditions the test ends with probability one and its stopping time is almost surely minimized in the class of tests with the  same or smaller error probabilities. Furthermore, reinforcing one of such conditions, it is also shown that any moment of the stopping time distribution is first-order asymptotically minimized in the same class of tests.
\item[-] At the application stage, the general problem of multi-frame target detection and tracking for radar surveillance is considered: in this way, previous limitations on target mobility imposed by other studies are avoided.
\item[-] Finally, a thorough performance analysis is given, aimed primarily at showing the correctness of the asymptotic analysis and at investigating the effects of system parameters. The superiority of sequential detection and estimation rules with respect to FSS techniques is also shown in the afore-mentioned radar application.
\end{itemize}

The rest of the paper is organized as follows. Next section presents the elements of the problem while section \ref{seq_rule_section} addresses the sequential detection and estimation problem. Section \ref{asym_an_section} presents the asymptotic results while section \ref{radar_appl_sec} covers the radar surveillance problem. Finally, section \ref{num_res_sec} is devoted to the presentation of numerical results, while concluding remarks are given in section \ref{conclusions}. For reader's sake, some notation, used throughout the rest of the paper, is first introduced.

\paragraph*{Notation} In what follows, all random variables are defined on a common probability space $(\Omega, \mathscr{F}, P)$ and are denoted with capital letters. Lower case letters are used to denote realizations of random variables while calligraphic letters to denote sets within which random variables take values. $\sigma$-algebras are denoted using script letters, $\sigma(X)$ being the smallest $\sigma$-algebra generated by the random variable $X$. $\mathbf{X}_{i:j}$ will be used to denote segments of random variables taken from the process $\{X_k\}_{k\in\mathbb{Z}}$: specifically, $\mathbf{X}_{i:j}=\{X_k\}_{k=i}^j$ for $i\leq j$, and $\mathbf{X}_{-\infty:j}= \{X_k\}_{k= -\infty}^j$. $\operatorname{E}$ is the operator of expectation: a subscript will be added in case of ambiguity, so that $\operatorname{E}_\theta$ and $\operatorname{E}_{H}$ are expectation when $\theta$ is the true state of nature and hypothesis $H$ is true, respectively. $\operatorname{D}(\cdot\Vert\cdot)$ denotes the Kullback-Leibler divergence operator. The acronyms a.s. and a.e. stands for almost sure and almost everywhere. $\mathbb{N}$ denotes the set of natural numbers, i.e. $\{1,2,\ldots\}$, $\mathbb{Z}$ the set of integers, $\mathbb{R}$ the set of real numbers and $\mathbb{R}^+$ the set of positive real numbers. Finally, the notation $h_v\sim g_v$ means that $\lim\limits_{v \rightarrow 0}h_v/g_v=1$.

\section{Problem formulation}\label{prob_form_sec}
Consider a dynamic system with a Markov evolution. $X_i$, $i\in\mathbb{N}$, is the state vector at time $i$ and $\mathcal{S}$ is the state space, with cardinality $M$. In particular, $\{X_i\}_{i\in\mathbb{N}}$ forms a discrete-time, homogeneous Markov chain with given initial distribution $\pi$ and transition probabilities $a(x_i,x_j)=P \big(\{X_k=x_i\}| \{X_{k-1}=x_j\}\big)$, $x_i,x_j\in\mathcal{S}$. A sequence of states $\{X_i\}_{i=1}^k$, often called trajectory, is denoted with $\mathbf{X}_{1:k}$ and has density $p_k(\mathbf{x}_{1:k})=\pi(x_1) \prod_{i=2}^k a(x_{i-1}, x_i)$, with respect to the counting measure. $\{X_i\}_{i\in\mathbb{N}}$ is observed through a set of noisy measurements. The measurement process is $\{Z_i\}_{i\in\mathbb{N}}$, and the sample space of each $Z_i$ is $(\mathcal{Z},\mathscr{V})$, $\mathscr{V}$ being a $\sigma$-algebra of subsets of $\mathcal{Z}$. Consider a $\sigma$-finite measure $\nu$ on $(\mathcal{Z},\mathscr{V})$. If the signal $\{X_i\}_{i\in\mathbb{N}}$ is present, $\{(X_i, Z_i)\}_{i\in\mathbb{N}}$ is a HMM: given a realization $\{x_i\}_{i\in\mathbb{N}}$ of $\{X_i\}_{i\in\mathbb{N}}$, $\{Z_i\}_{i\in\mathbb{N}}$ is a sequence of conditionally independent random variables, each $Z_i$ having density $f(z|x_i)$ with respect to $\nu$. On the other hand, if the measurements contain only noise, $\{Z_i\}_{i\in\mathbb{N}}$ is an i.i.d. process, each $Z_i$ having density $f(z|\theta_0)$ with respect to $\nu$. Thus, for every $k\in\mathbb{N}$, the joint distribution of $\mathbf{Z}_{1:k}$ has conditional density
\begin{equation}
\begin{aligned}
f_k(\mathbf{z}_{1:k}|\mathbf{x}_{1:k})&=\prod_{i=1}^k f(z_i|x_i), && \begin{aligned} &\text{if the signal is present}\\ &\text{and } \mathbf{X}_{1:k}=\mathbf{x}_{1:k},\end{aligned}\\
f_k(\mathbf{z}_{1:k}|\theta_0)&=\prod_{i=1}^k f(z_i|\theta_0),&& \text{if the signal is not present},
\end{aligned}\notag
\end{equation}
with respect to $\nu^k$.

Given these elements, one is to sample the process $\{Z_i\}_{i\in\mathbb{N}}$ sequentially and decide, as soon as possible, if measurements are generated by noise alone or if they come from a dynamic system. In the latter case, it can be also required to estimate the system trajectory which has generated such measurements. The parameter space, then, is $\{\theta_0\}\cup\big( \mbox{\begin{Large}$\times$\end{Large}} _{i\in\mathbb{N}} \mathcal{S}\big)$. As in \cite{Mid1968, Fre1972, Bay1995}, whose focus, however, was on non-sequential decision rules, there is a mutual coupling of detection and estimation and two different strategies may be adopted. Indeed, the structure of the decision rule can be chosen so as to improve the detection or the estimation performance. The former case is called a weakly coupled (or uncoupled) design while the latter a strongly coupled (or coupled) design. In both cases, the estimator is enabled by the detection operation: this gating, however, can be (possibly) optimal for the detection or for the estimation. However,
the problem of designing sequential procedures for detection and estimation is considerably more difficult than that of devising FSS procedures \cite{Bla1954} and the approach taken in general is to extend and generalize the SPRT designing a practical, possibly sub-optimal, rule \cite{Arm1950, Mar1962, Lor1972, Lor1977, Tar1999, Dra1999, Dra2000}. In this paper the uncoupled strategy is adopted, this choice being motivated by a number of reasons: it has a very simple structure; as shown in section \ref{asym_an_section}, it exhibits many optimal properties; detection is the primary interest in many practical applications, as for example, radar surveillance problems later discussed.

\section{Detection and estimation procedure}\label{seq_rule_section}

A sequential decision rule is the pair $(\varphi,\xi)$, where $\varphi=\{\varphi_k \}_{k\in\mathbb{N}}$ is a stopping rule and $\xi=\{\xi_k \}_{k\in\mathbb{N}}$ a terminal decision rule \cite{Fer1967}. Since detection and estimation are performed in parallel, the terminal decision rule is itself composed of a detection rule $\delta=\{\delta_k \}_{k\in\mathbb{N}}$ for testing the signal presence and of a trajectory estimator $\widehat{x} = \{\widehat{x}_k \}_{k\in\mathbb{N}}$, i.e. $\xi=(\delta,\widehat{x})$. The proposed (non-randomized) sequential decision rule is, then,
\begin{subequations}\label{SPRT_rule}
\begin{align}
\varphi_k(\mathbf{z}_{1:k})=&\begin{cases}1,&\text{if } \Lambda_k(\mathbf{z}_{1:k}) \notin
(\gamma_0,\gamma_1),\\0,& \text{otherwise},
\end{cases}\\
\delta_k(\mathbf{z}_{1:k})=&\begin{cases}1,&\text{if } \Lambda_k(\mathbf{z}_{1:k}) \geq\gamma_1,\\ 0,& \text{if } \Lambda_k(\mathbf{z}_{1:k}) \leq\gamma_0,
\end{cases}\\
\widehat{x}_k(\mathbf{z}_{1:k})=&\operatorname*{arg\,max}_{\mathbf{x}_{1:k}\in\mathcal{S}^k} p_k(\mathbf{x}_{1:k}) \Lambda_k(\mathbf{z}_{1:k}|\mathbf{x}_{1:k}),\text{ if } \Lambda_k(\mathbf{z}_{1:k}) \geq\gamma_1,\label{estimation_rule}
\end{align}
\end{subequations}
where $\Lambda_k(\mathbf{z}_{1:k})= \sum_{\mathbf{x}_{1:k}\in\mathcal{S}^k} p_k(\mathbf{x}_{1:k}) \Lambda_k(\mathbf{z}_{1:k}| \mathbf{x}_{1:k})$,  $\Lambda_k(\mathbf{z}_{1:k}|\mathbf{x}_{1:k})$ being the likelihood ratio of $f_k(\mathbf{z}_{1:k}|\mathbf{x}_{1:k})$ to $f_k(\mathbf{z}_{1:k}| \theta_0)$.

Notice that the pair $(\varphi,\delta)$ is an SPRT for testing $H_0=$ `noise only' against the alternative $H_1=$ `signal present', no matter of its trajectory. $H_1$, then, is the hypothesis that $\mathbf{Z}_{1:k}$ has density $f_{k,H_1}(\mathbf{z}_{1:k})= \sum_{\mathbf{x}_{1:k}\in \mathcal{S}^k}p_k(\mathbf{x}_{1:k})f_k (\mathbf{z}_{1:k}| \mathbf{x}_{1:k})$, $\forall\;k\in\mathbb{N}$. The strength of such a sequential test is the pair of probabilities of errors of the first and second kind, $\alpha$ and $\beta$, respectively (often, in detection problems, $\alpha$ is referred to as probability of false alarm, $P_\text{fa}$, and $\beta$ as probability of miss, $P_\text{miss}$). Denoting with $\tau$ the stopping time and with $\psi=\{\psi_k\}_{k\in\mathbb{N}}$ its conditional distribution,\footnote{$\psi_k( \mathbf{z}_{1:k})$ is the probability that $\tau=k$ given a realization $\mathbf{z}_{1:k}$ of $\mathbf{Z}_{1:k}$, for any $k\in\mathbb{N}$; the relationship between $\psi$ and $\varphi$ is: $\psi_1(z_1)=\varphi_1(z_1)$ and $\psi_k(\mathbf{z}_{1:k})= \varphi(\mathbf{z}_{1:k}) \prod_{\ell=1}^{k-1}\big(1-\varphi(\mathbf{z}_{1:\ell})\big)$, for $k>1$.} these probabilities of error are given by $\alpha= \sum_{k\in\mathbb{N}} \operatorname{E}_{\theta_0} \big[\psi_k(\mathbf{Z}_{1:k}) \delta_k(\mathbf{Z}_{1:k})\big]$ and $\beta=\sum_{k\in\mathbb{N}}\operatorname{E}_{H_1} \big[\psi_k(\mathbf{Z}_{1:k}) \big(1-\delta_k(\mathbf{Z}_{1:k})\big)\big]$, $P_\text{d}=\sum_{k\in\mathbb{N}} \operatorname{E}_{H_1} \big[\psi_k(\mathbf{Z}_{1:k}) \delta_k(\mathbf{Z}_{1:k})\big]$ being the probability of detection.\footnote{Notice that $\beta+P_\text{d}= P\big(\{\tau<+\infty\}|H_1\big)$.} The boundaries of the test, $\gamma_0$ and $\gamma_1$, with $0<\gamma_0<1< \gamma_1<+\infty$, are chosen in order to have the required strength $(\alpha,\beta)$.

As concerns $\widehat{x}$ only, it can be considered a gated estimator since estimation is enabled by the detection rule. Furthermore, consider the triplet $(\tau, \mathbf{X}_{1:\tau},\mathbf{Z}_{1:\tau})$. Since
\begin{align}
&P\big( \{  \tau=k, \mathbf{X}_{1:\tau} =\mathbf{x}_{1:k} , \mathbf{Z}_{1:\tau}  \in A_k\} | \{\tau<+\infty, \notag\\
& \delta_\tau(\mathbf{Z}_{1:\tau})=1,H_1\} \big) =\notag\\
&=\frac{P\big( \{  \tau=k, \mathbf{X}_{1:\tau} =\mathbf{x}_{1:k} , \mathbf{Z}_{1:\tau}  \in A_k,\delta_\tau(\mathbf{Z}_{1:\tau})=1\}| H_1 \big)}{P\big( \{ \tau<+\infty,\delta_\tau(\mathbf{Z}_{1:\tau})=1\}| H_1 \big)}=\notag\\
&= P_\text{d}^{-1} \int_{A_k} p(\mathbf{x}_{1:k}) f_k(\mathbf{z}_{1:k}| \mathbf{x}_{1:k}) \psi_k(\mathbf{z}_{1:k}) \delta_k(\mathbf{z}_{1:k}) d\nu^k(\mathbf{z}_{1:k}) ,\notag
\end{align}
$\forall\;k\in\mathbb{N}$, $\mathbf{x}_{1:k}\in\mathcal{S}^k$ and $A_k\in\sigma(\mathbf{Z}_{1:k})$, $P_\text{d}^{-1} p(\mathbf{x}_{1:k}) f(\mathbf{z}_{1:k}| \mathbf{x}_{1:k}) \psi_k(\mathbf{z}_{1:k}) \delta_k(\mathbf{z}_{1:k})$ is the density of $(\tau, \mathbf{X}_{1:\tau},\mathbf{Z}_{1:\tau})|\{\text{accept}\;H_1, H_1\; \text{true}\}$. This means that $P_\text{d}^{-1} \sum_{k\in\mathbb{N}}p(\mathbf{x}_{1:k}) f(\mathbf{z}_{1:k}| \mathbf{x}_{1:k}) \psi_k(\mathbf{z}_{1:k}) \delta_k(\mathbf{z}_{1:k})$ is the density of $(\mathbf{X}_{1:\tau},\mathbf{Z}_{1:\tau})|\{\text{accept }H_1, H_1 \text{ true}\}$ so that
\begin{equation}
\operatorname*{arg\;max}_{\theta\in\mbox{\begin{normalsize}$\times$\end{normalsize}} _{i\in\mathbb{N}} \mathcal{S}} \sum_{k\in\mathbb{N}}  p(\mathbf{x}_{1:k}) f(\mathbf{Z}_{1:k}| \mathbf{x}_{1:k}) \psi_k(\mathbf{Z}_{1:k}) \delta_k(\mathbf{Z}_{1:k})\notag
\end{equation}
is a MAP estimator conditioned upon the event $\{\text{accept }H_1, H_1 \text{ true}\}$, $\mathbf{x}_{1:k}$ being the projection of $\theta$ on $\mathcal{S}^k$. Since the above estimation rule is exactly that in (\ref{estimation_rule}), it results that $\widehat{x}$ is a MAP estimation rule conditioned that no error of the first kind is made by the detector.

Finally, as to the computational complexity, the sequential decision rule in (\ref{SPRT_rule}) requires to evaluate the statistics $\sum_{\mathbf{x}_{1:k} \in\mathcal{S}^k} p_k(\mathbf{x}_{1:k}) \Lambda_k(\mathbf{Z}_{1:k}| \mathbf{x}_{1:k})$ and $\max_{\mathbf{x}_{1:k}\in\mathcal{S}^k} p_k(\mathbf{x}_{1:k}) \Lambda_k(\mathbf{Z}_{1:k}| \mathbf{x}_{1:k})$ for $k=1,\ldots,\tau$, where $p_k(\mathbf{x}_{1:k}) \Lambda_k(\mathbf{Z}_{1:k})= \pi(x_1)\frac{f(Z_1| x_1)}{f(Z_1|\theta_0)} \prod_{i=2}^k a(x_{i-1},x_i)\frac{f(Z_i|x_i)}{f(Z_i|\theta_0)}$. These statistics, then, have the form of a stage-separated function on the algebraic system $(\mathbb{R},+,\cdot)$ and $(\mathbb{R}, \max, \cdot)$, respectively, and can be computed through two dynamic programming algorithms \cite{Ver1987} (similar to the forward-backward procedure and to the Viterbi algorithm as in \cite{Gro2008}), whereby the computational complexity is only linear in $k$. Notice that maximization in (\ref{estimation_rule}) is preferred to $\arg \max_{\mathbf{x}_{1:k}\in \mathcal{S}^k} p_k(\mathbf{x}_{1:k}) f_k(\mathbf{z}_{1:k}| \mathbf{x}_{1:k})$ since, in the former case, the estimator can work on the same data as the detector, thus further lowering the complexity.

\section{Asymptotic analysis}\label{asym_an_section}

Let $\mathscr{T}(\alpha',\beta')$, $\alpha',\beta'\in(0,1)$, be the class of non-randomized tests (denoted with $(N,d)$, $N$ being the stopping time and $d$ the terminal decision rule), either sequential or FSS, with probability of error of the first and second kind bounded by $\alpha'$ and $\beta'$, respectively. It is known that Wald's SPRT for testing a simple hypothesis against a simple alternative based on i.i.d. observations has the following remarkable properties  \cite{Wal1945, Wal1948, Bur1963, Fer1967}.
\begin{enumrm}
\item\label{SPRTprop1} If the test has strength $(\alpha,\beta)$ and boundaries $\gamma_0$, $\gamma_1$, then
\begin{equation}
\alpha\leq (1-\beta)/\gamma_1\leq 1/\gamma_1 \quad \text{and} \quad \beta \leq (1-\alpha)\gamma_0\leq \gamma_0.\label{approx_bound}
\end{equation}
\item\label{SPRTprop2} The test ends a.s. under both hypotheses.
\item\label{SPRTprop3} The ASN is finite under both hypotheses.
\item\label{SPRTprop4} The ASN is minimized among tests in the class $\mathscr{T}(\alpha',\beta')$ under both hypotheses.
\end{enumrm}
Except property (\ref{SPRTprop1}), which is easily shown to hold under very general conditions \cite{Ber1973}, the other properties in general do not hold in the present setting since the observations $\{Z_i\}_{i\in\mathbb{N}}$ are not independent. This section is devoted to studying the asymptotic behaviour of the sequential test when the two error probabilities simultaneously approach zero. It will be demonstrated that, under rather mild hypotheses, the procedure satisfies also properties (\ref{SPRTprop2}) and (\ref{SPRTprop3}) and, asymptotically, (\ref{SPRTprop4}). In particular, it will be shown that every finite moment of the stopping time is first-order asymptotically minimized in the class $\mathscr{T}(\alpha',\beta')$. The regularity conditions are stated below.
\begin{condition}\label{asym_cond_1}
The Markov chain $\{X_k\}_{k\in\mathbb{N}}$ is stationary, irreducible and aperiodic.
\end{condition}
\begin{condition}\label{asym_cond_2}
The family of mixtures of at most $M$ elements of
$\{f(\cdot|x)\}_{x\in\mathcal{S}}$ is not equal to $f(\cdot|\theta_0)$ $\nu$-a.e., i.e.,
for every distribution $c$ on $\mathcal{S}$, $\sum_{x\in\mathcal{S}}c(x) f(\cdot|x)$ and $f(\cdot|\theta_0)$ are equal $\nu$-a.e..
\end{condition}
\begin{condition}\label{asym_cond_3}
For every $x\in\mathcal{S}$, $\operatorname{E}_{H_i} \left[ \left| \ln \frac{f(Z_1|x)}{f(Z_1|\theta_0)}\right| \right]<+\infty$, $i=0,1$.
\end{condition}
\begin{condition}\label{asym_cond_4}
There exists a constant $a>0$ such that $\operatorname{E}_{H_0}\Big[\left( \frac{f(Z_1|x)}{f(Z_1|\theta_0)} \right)^{\pm a}\Big]$ and $\operatorname{E}_y\Big[\left( \frac{f(Z_1|x)}{f(Z_1|\theta_0)} \right)^{\pm a}\Big]$ are finite, for all $x,y\in\mathcal{S}$.
\end{condition}
\begin{condition}\label{asym_cond_5}
$\left(\frac{f(z|x)}{f(z|\theta_0)}\right)^{\pm1}\neq 0$ for every $x\in\mathcal{S}$ and $z\in\mathcal{Z}$.
\end{condition}
\begin{condition}\label{asym_cond_6}
The matrix containing the transition probabilities $\{a(x,y)\}_{x,y\in\mathcal{S}}$ is invertible.
\end{condition}

\begin{remark}
Since the Markov chain $\{X_i\}_{i\in\mathbb{N}}$ is homogeneous and has a finite state space, condition \ref{asym_cond_1} corresponds to requiring that $\{X_i\}_{i\in\mathbb{N}}$ be stationary and ergodic, which will be seen to imply $\{Z_i\}_{i\in\mathbb{N}}$ to be stationary and ergodic as well, an essential property for the limiting theorem to be presented. As concerns condition \ref{asym_cond_2}, it can be shown (recursively) that it ensures the two densities $f_k (\cdot| \theta_0)$ and $f_{k,H_1}(\cdot)$ are not $\nu$-a.e. equal for every $k\in\mathbb{N}$: otherwise, for some $k\in\mathbb{N}$ it could not be possible to discriminate between statistical populations drawn from these two distributions, i.e. detection could not be possible. Finally, \ref{asym_cond_3} -- \ref{asym_cond_6} are essentially `regularity' conditions which allow to derive the limiting behaviour of the log-likelihood ratios $\{\ln\Lambda_k\}_{k\in\mathbb{N}}$. Notice, furthermore, that the moment conditions \ref{asym_cond_4} imply \ref{asym_cond_3} since there always exists a finite constant $C$ such that $|\ln w|\leq 2C \cosh(a\ln w)= C\left(w^a + w^{-a}\right)$, for any $w>0$.
\end{remark}

It turns out that the validity of properties (\ref{SPRTprop2}) -- (\ref{SPRTprop4}) is highly influenced by the limits $\lambda_i= \lim\limits_{k\rightarrow+ \infty}k^{-1} \operatorname{E}_{H_i}\big[\ln \Lambda_k(\mathbf{Z}_{1:k})\big]$, $i=0,1$, in the case that they exist finite and non-zero \cite{Lai1981}. For this reason, it is first given the following (proof is given in the Appendix).
\begin{theorem}\label{theorem_conv_as}
If conditions \ref{asym_cond_1} -- \ref{asym_cond_3} are fulfilled, there exist
finite constants $\lambda_0<0$ and $\lambda_1>0$ such that
\begin{align}
&\lim_{k\rightarrow+\infty}\frac{1}{k} \operatorname{E}_{H_i}\big[\ln \Lambda_k(\mathbf{Z}_{1:k})\big]=\lambda_i,\notag\\
&\lim_{k\rightarrow+\infty}\frac{1}{k} \ln \Lambda_k(\mathbf{Z}_{1:k})
=\lambda_i,\text{ a.s. under } H_i,\quad i=0,1.\notag
\end{align}
These conclusions hold for any initial probability with strictly positive entries (i.e. not necessarily the stationary one) used in the definition of $\{\Lambda_k\}_{k\in\mathbb{N}}$ and $\lambda_i$ have the same value.
\end{theorem}
If the log-likelihood ratios $\{\ln \Lambda_k(\mathbf{Z}_{1:k})\}_{k\in\mathbb{N}}$ satisfy this sort of `stability' property, it can be easily demonstrated the following
\begin{proposition}\label{test_end_as}
Under conditions \ref{asym_cond_1} -- \ref{asym_cond_3}, the test ends a.s. under both hypotheses, i.e. $P\big(\{\tau<+\infty\}|H_i\big)=1$, $i=1,0$.
\end{proposition}
\begin{IEEEproof}
Define the two auxiliary stopping times $\tau_0=\inf\big\{ k\in\mathbb{N}:\; \Lambda_k(\mathbf{Z}_{1:k})\leq \gamma_0 \big\}$ and $\tau_1=\inf\big\{ k\in\mathbb{N}:\; \Lambda_k(\mathbf{Z}_{1:k})\geq \gamma_1 \big\}$. From theorem \ref{theorem_conv_as}, $\lim_{k\rightarrow+\infty}k^{-1} \ln \Lambda_k(\mathbf{Z}_{1:k})=\lambda_i$, a.s. under $H_i$, $i=0,1$, with $\lambda_0<0$ and $\lambda_1>0$. This implies that
\begin{align}
P\Big(\Big\{ \lim_{k\rightarrow+\infty} \Lambda_k(\mathbf{Z}_{1:k})<\gamma_0\Big\}|H_0\Big)&=1, \notag\\
P\Big(\Big\{ \lim_{k\rightarrow+\infty} \Lambda_k(\mathbf{Z}_{1:k})>\gamma_1\Big\}|H_1\Big)&=1,\notag
\end{align}
which means that $P\big(\{\tau_i<+\infty\}|H_i\big)=1$, $i=0,1$. The thesis, then, follows form the fact that $\tau=\min\{\tau_0,\tau_1\}$.
\end{IEEEproof}
The `stability' of $\{\ln \Lambda_k(\mathbf{Z}_{1:k})\}_{k\in\mathbb{N}}$ of theorem \ref{theorem_conv_as} is sufficient also to imply the asymptotic optimality of the test in the sense of the following
\begin{theorem}\label{asym_suboptimality}
Suppose that conditions \ref{asym_cond_1} -- \ref{asym_cond_3} are fulfilled, $\gamma_0$ and $\gamma_1$ are chosen so that the test belong to $\mathscr{T}(\alpha',\beta')$ and $\ln\gamma_1\sim \ln\frac{1}{\alpha'}$, $\ln\gamma_0 \sim\ln \beta'$ as $\alpha'+\beta'\rightarrow0$. Then
\begin{align}
&\frac{\tau}{|\ln\beta'|}\xrightarrow[\alpha'+\beta'\rightarrow0]{}\frac{1}{|\lambda_0|}, && \text{a.s. under }H_0,\notag\\
&\frac{\tau}{|\ln \alpha'|}\xrightarrow[\alpha' +\beta'\rightarrow0]{} \frac{1}{\lambda_1}, && \text{a.s. under }H_1.\notag
\end{align}
Furthermore, for every $\varepsilon\in(0,1)$, it results
\begin{equation}
\inf_{(N,d)\in\mathscr{T}(\alpha',\beta')}P\big(\big\{N>\varepsilon\tau\big\}|H_i
\big)\xrightarrow[\alpha'+\beta'\rightarrow0]{}1,\quad
i=0,1.\label{quasiopt_prop}
\end{equation}
\end{theorem}
\begin{IEEEproof}
Under conditions \ref{asym_cond_1} -- \ref{asym_cond_3}, the conclusions of theorem \ref{theorem_conv_as} hold and then theorem 1 of \cite{Lai1981} can be used.
\end{IEEEproof}
Notice that equation (\ref{approx_bound}) allows to choose the appropriate thresholds: indeed, it implies that  $\gamma_1=1/\alpha'$ and $\gamma_0=\beta'$ result in a test which belongs to $\mathscr{T}(\alpha',\beta')$. Moreover, regarding (\ref{approx_bound}) as approximate equalities (i.e. by neglecting overshoots) leads to the approximations (see \cite{Wal1945})
\begin{equation}
\gamma_1\approx(1-\beta)/\alpha\overset{\alpha+\beta\rightarrow0}
\sim 1/\alpha,\quad \gamma_0\approx \beta/(1-\alpha)
\overset{\alpha+\beta\rightarrow0} \sim \beta.\label{boundrs_sett}
\end{equation}
Notice that (\ref{quasiopt_prop}) does not imply asymptotic optimality of the test (the optimality criterion of the Wald-Wolfwitz theorem \cite{Wal1948} is about the minimization of the ASN under both hypotheses). With the a.s. convergence of $\{k^{-1}\ln\Lambda_k (\mathbf{Z}_{1:k})\}_{k\in\mathbb{N}}$ alone, the following can be proved.
\begin{theorem}\label{thm_lowbounds}
If conditions \ref{asym_cond_1} -- \ref{asym_cond_3} are satisfied then
\begin{equation}
\liminf_{\alpha'+\beta'\rightarrow0}\frac{\operatorname{E}_{H_0}[\tau^r]}{|\ln\beta'|^r}\geq\frac{1}{|\lambda_0|^r} \quad \text{and} \quad \liminf_{\alpha'+\beta'\rightarrow0} \frac{\operatorname{E}_{H_1}[\tau^r]}{|\ln\alpha'|^r} \geq \frac{1}{ \lambda_1^r},\notag
\end{equation}
for every positive constant $r$.
\end{theorem}
\begin{IEEEproof}
As in \cite[theorem 2.2]{Tar1999}, it is sufficient to apply the Markov's inequality. Indeed, it results
\begin{equation}
\operatorname{E}_{H_0}\!\bigg[ \!\left(\frac{\tau|\lambda_0|}{|\ln\beta'|} \right)^r\!\bigg]\geq \varepsilon^r P\left(\!\left\{ \frac{\tau|\lambda_0|}{|\ln\beta'|}>\varepsilon \big|H_0 \right\}\!\right)\! \xrightarrow[\alpha'+\beta'\rightarrow0]{} \varepsilon^r\!, \notag
\end{equation}
for any $r>0$ and $\varepsilon\in (0,1)$, where the limit follows from theorem \ref{asym_suboptimality}. Since $\varepsilon$ is an arbitrary constant in $(0,1)$, it follows that $\liminf_{\alpha'+\beta'\rightarrow0} \operatorname{E}_{H_0}\left[ \left(\frac{\tau|\lambda_0|}{|\ln\beta'|} \right)^r\right]\geq 1$, which proves the first inequality. The other one can be proved similarly.
\end{IEEEproof}
In order to guarantee finiteness of the expected sample size and to obtain its first-order asymptotic minimization, condition \ref{asym_cond_3} must be strengthened requiring \ref{asym_cond_4} -- \ref{asym_cond_6} to hold. Indeed, the following can be proved (proof is given in the Appendix).
\begin{theorem}\label{asym_optimality}
Suppose that conditions \ref{asym_cond_1}, \ref{asym_cond_2}, \ref{asym_cond_4} -- \ref{asym_cond_6} are fulfilled, $\gamma_0$ and $\gamma_1$ are chosen so that the test belongs to $\mathscr{T}(\alpha',\beta')$ and $\ln\gamma_1\sim \ln\frac{1}{\alpha'}$, $\ln\gamma_0 \sim\ln \beta'$ as $\alpha'+\beta'\rightarrow0$. Then, for every $r\in\mathbb{N}$, $\operatorname{E}_{H_i}[\tau^r]<+\infty$, $i=0,1$, and, as $\alpha'+\beta'\rightarrow0$,
\begin{align}
\inf_{(N,d)\in\mathscr{T}(\alpha',\beta')}\operatorname{E}_{H_0}[N^r] \sim \operatorname{E}_{H_0}[\tau^r] \sim \frac{|\ln\beta'|^r}{|\lambda_0|^r},\notag\\
\inf_{(N,d)\in\mathscr{T}(\alpha',\beta')}\operatorname{E}_{H_1}[N^r] \sim \operatorname{E}_{H_1}[\tau^r] \sim \frac{|\ln\alpha'|^r}{\lambda_1^r}.\notag
\end{align}
\end{theorem}
The $\mathscr{O}(1)$ term is due to the overshoot $\ln\big(\Lambda_\tau (\mathbf{Z}_{1:\tau})/ \gamma_0\big)$ or $\ln\big(\Lambda_\tau(\mathbf{Z}_{1:\tau})/ \gamma_1\big)$, analogous to Wald's lower bound \cite{Wal1945} for the sample size (which is attained when these overshoots are ignored).

From theorems \ref{asym_suboptimality} and \ref{asym_optimality}, it results that the asymptotic behaviour of the test is determined by the constants $\lambda_i$, $i=0,1$. These constants are often difficult to evaluate and, thus, approximations or bounds can be useful. To this end, the following propositions, whose proofs are given in the Appendix, are presented.
\begin{proposition}\label{prop_bound_sup}
Constants $\lambda_i$, $i=0,1$, satisfy the following
\begin{subequations}
\begin{align}
\lambda_1 & \leq \sum_{x\in\mathcal{S}}\overline{\pi}(x)D\big(f(\cdot|x)\Vert f(\cdot|\theta_0) \big) ,\\
|\lambda_0|& \leq \sum_{x\in\mathcal{S}}\overline{\pi}(x)D\big(f(\cdot|\theta_0)\Vert f(\cdot|x)\big),
\end{align} \label{temp1}
\end{subequations}
where $\overline{\pi}$ is the unique stationary distribution of the Markov chain $\{X_i\}_{i\in\mathbb{N}}$.
\end{proposition}

\begin{proposition}\label{prop_bounds}
If for every distribution $c$ on $\mathcal{S}$
\begin{subequations}
\begin{align}
\textstyle \operatorname{D}\Big(\sum\limits_{x\in\mathcal{S}}c(x)f(\cdot|x) \Vert f(\cdot|\theta_0)\Big)&\textstyle =\operatorname{D}\Big(\sum\limits_{x\in\mathcal{S}}c'(x)f(\cdot|x)\big\Vert f(\cdot|\theta_0)\Big),\\
\textstyle \operatorname{D}\Big( f(\cdot|\theta_0)\Vert \sum\limits_{x\in\mathcal{S}}c(x)f(\cdot|x)\Big)&\textstyle =\operatorname{D}\Big(f(\cdot|\theta_0)\Vert \sum\limits_{x\in\mathcal{S}}c'(x)f(\cdot|x)\Big),
\end{align}\label{perm_condition}
\end{subequations}
for any permutation $c'$ of $c$, then
\begin{subequations}
\begin{gather}
\textstyle D\Big(\frac{1}{M}\sum\limits_{y\in\mathcal{S}} f(\cdot|y) \Vert f(\cdot|\theta_0)\Big) \leq \lambda_1 \leq D\big(f(\cdot|x)\; \Vert\; f(\cdot|\theta_0)\big),\\
\textstyle D\Big(f(\cdot|\theta_0)\Vert \frac{1}{M}\sum\limits_{y\in\mathcal{S}} f_1(\cdot|y)\Big) \leq |\lambda_0| \leq D\big(f(\label{bounds_on_lambda}\cdot|\theta_0)\;\Vert\; f(\cdot|x)\big),
\end{gather}
\end{subequations}
where $D\big(f(\cdot|x)\; \Vert\; f(\cdot|\theta_0)\big)$ and $D\big(f(\cdot|\theta_0)\;\Vert\; f(\cdot|x)\big)$ assume the same value for any $x\in\mathcal{S}$.
\end{proposition}
Notice that the upper bounds in (\ref{bounds_on_lambda}) are attained if $\pi(x)=1$ for some $x\in\mathcal{S}$ and $a(x,x)=1$, $\forall\, x\in\mathcal{S}$, while the lower bounds if $\pi(x)=1/M$, $\forall\, x\in\mathcal{S}$, and $a(x,y)=1/M$, $\forall x,y\in\mathcal{S}$.

\section{Example of application: the radar case}\label{radar_appl_sec}

The radar problem is characterized by the inherent presence of multiple-resolution elements, which correspond to range `bins' as well as Doppler, azimuth and elevation cells. This problem has been solved in \cite{Mar1962, Hel1962, Tar2003} but all of these approaches concern the case that the target is not allowed to change its position while being illuminated by the radar. This condition may be too restrictive, especially in airborne applications where the relative radial velocity between target and radar may exceed Mach-2. The surveillance area is divided into smaller angular regions, each visited in turn by the antenna beam in cyclic manner. In each region a sequential procedure is used to accept or reject the hypothesis that a single target is present. The measurement process is obtained dividing the region into a grid and discretizing the continuous-time received signal accordingly (if the grid is sufficiently fine, losses due to possible mismatches may be neglected). The measurement at epoch $i\in\mathbb{N}$, also called frame, is the set of returns received from all of the radar resolution elements, i.e. $Z_\ell=\big\{Z_\ell(x): x\in\{1,\ldots,N_a\} \times\{1,\ldots,N_e\}\times\{1, \ldots,N_r\}\times\{1,\ldots,N_d\} \big\}$, where $N_a, N_e,N_r,N_d$ are the number of resolution elements in azimuth, elevation, range and Doppler, respectively. The target signature appears on at most one resolution element in each frame. The target state space consists of the set of all the resolution cells, i.e. $\mathcal{S}=\{1,\ldots,N_a\} \times \{1,\ldots,N_e\} \times\{1,\ldots,N_r\}\times \{1,\ldots,N_d\}$, with $M=N_aN_eN_rN_d$. If also velocities are to be considered, then the state space can be enlarged consequently. A first-order Gaussian-Markov random walk model is used to derive the transition probabilities, which are given by $a(x_i, x_{i+1})=\prod_{\ell=1}^4 a_\ell(x_{i,\ell}, x_{i+1,\ell})$, where $x_{i,\ell}$ denotes the $\ell$-th component of the target state vector at epoch $i$ and
\begin{align}
a_\ell(x_{i,\ell}, x_{i+1,\ell})=& Q\Big(\frac{x_{i+1,\ell} -x_{i,\ell}-1/2}{\sigma_\ell}\Big)+ \notag\\
&-Q\Big(\frac{x_{i+1,\ell}-x_{i,\ell}+1/2}{\sigma_\ell} \Big),\quad\ell=1,\ldots,4. \notag
\end{align}
In the above equation, $Q(x)=\frac{1}{\sqrt{2\pi}}\int_x^{+\infty}e^{-t^2}dt$ and $\sigma_\ell$ is a parameter related to the target mobility along the $\ell$-th dimension: large values of $\sigma_\ell$ allow large target maneuvers but decrease, at the same time, target detection and estimation capabilities.\footnote{Notice that, even if all transitions are theoretically admissible, real targets necessarily need to satisfy physical constraints, such as limitations on the maximum velocity and acceleration. In this case, a truncated Gaussian density can be used.} As concerns the initial probability, if no other prior information is available (for example previous detections), it is reasonable to force $\pi(x)=1/M$, for all $x\in\mathcal{S}$.

It is supposed that the components of the measurement $Z_i$ are independent, each $Z_i(x)$, $x\in\mathcal{S}$, being an exponentially distributed random variable with density
\begin{subequations}\label{densities_10}
\begin{align}
h_1(v)=&\frac{e^{-\frac{v}{1+\rho}}}{1+\rho}\operatorname{u}(v),&&\text{if the target is present in location $x$},\\
h_0(v)=&e^{-v}\operatorname{u}(v),&&\text{otherwise},
\end{align}
\end{subequations}
where $\rho$ denotes the signal-to-noise ratio (SNR) and $\operatorname{u}(y)$ is the unit-step function. In this case $f(z_i|x_i)=h_1\big(z_i(x_i)\big) \prod_{\genfrac{}{}{0pt}{3} {x\in\mathcal{S}\hfill}{x\neq x_i}} h_0\big(z_i(x)\big)$, $f(z_i|\theta_0)=\prod_{x\in\mathcal{S}} h_0\big(z_i(x)\big)$ and
\begin{equation}
\Lambda_k(\mathbf{z}_{1:k}) =\prod_{i=1}^k \frac{h_1\big(z_i(x_i)\big)}{h_0\big(z_i(x_i) \big)}=\prod_{i=1} ^k\frac{e^{z_i(x_i)\rho/ (1+\rho)}}{1+\rho}, \quad \forall \;k\in\mathbb{N}.\label{LR}\footnote{The model of equations \ref{densities_10} -- \ref{LR} applies, for examples, if measurements come from a square law envelope detector, the noise is additive, white and Gaussian, the target has a Swerling-I fluctuation model and frequency agility is used to achieve frame-to-frame target amplitude decorrelation. This is a common situation in radar scenarios \cite{Sko2001}.}
\end{equation}
Notice that, as it can be easily checked, this model satisfies conditions \ref{asym_cond_2} -- \ref{asym_cond_6}, and, also, equations (\ref{perm_condition}). In particular, for condition \ref{asym_cond_4}  to hold, it is sufficient to choose $a< 1/\rho$. This means that the test first-order asymptotically minimizes any positive moment of the stopping time distribution. Even in the above situation, however, occasionally long observations can be needed. Furthermore, if there are mismatches between design and actual values of some parameters (for example, the SNR) the resulting ASN can be very large, especially for small error probabilities. Truncation of the procedure then can be used to prevent such a problem: when a fixed sample $K$ is reached, hypothesis $H_1$ or $H_0$ is accepted whether $\Lambda_K(\mathbf{Z}_{1:k})$ exceeds $\gamma_K$ or not, respectively. The impact of truncation on the system performances as well as the problem of the final threshold setting is not explored further here and the reader is referred to the past literature \cite{bus1967, Tan1982,Tan1989}.

\section{Numerical results}\label{num_res_sec}

The behaviour of the sequential procedure has been tested through Monte Carlo simulations in terms of ASN and $P_\text{track}$, the probability of correct track estimation. First, a general problem of detection and trajectory estimation is considered in order to reinforce the discussion in sections \ref{seq_rule_section} and \ref{asym_an_section}. For simplicity, the measurement model is that of equations (\ref{densities_10}) -- (\ref{LR}), even if there is no explicit reference to a radar scenario. The state space is $\mathcal{S}=\{1,\ldots,M\}$ and the transition probabilities are derived from a truncated Gaussian distribution with standard deviation $\sigma$ using a quantization step of $10^{-4}$. The boundaries $\gamma_0$ and $\gamma_1$ have been set using equation (\ref{boundrs_sett}), where the design error probabilities $\alpha'=\beta'=10^{-3}$ have been adopted.
\begin{figure}[t]
\centering
\includegraphics[width=\columnwidth]{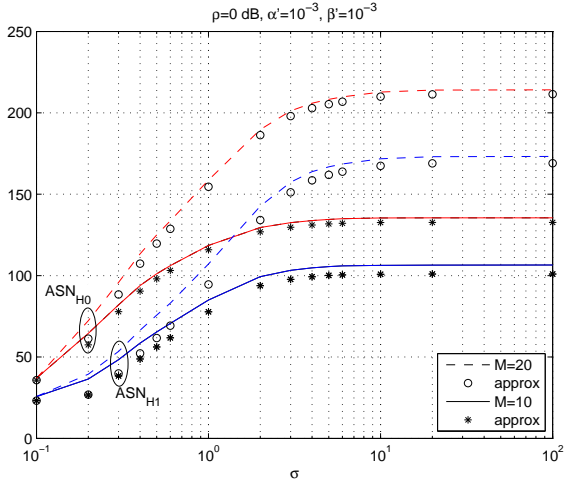}
\vspace{-0.7cm}\caption{ASN under both hypotheses versus the mobility parameter $\sigma$ for $\rho=0$ dB and different values of $M$. The markers denote the values resulting from the asymptotic approximations of the ASN of theorem \ref{asym_optimality}.} \label{fig1}
\end{figure}
In figure \ref{fig1} the ASN under both $H_0$ and $H_1$ is plotted versus $\sigma$ for $\rho=0$ dB and for various $M$. Since the model satisfies conditions \ref{asym_cond_1} -- \ref{asym_cond_6}, the approximations for the ASN of theorem \ref{asym_optimality} hold: the difference between the approximations and the true ASN is due to the excesses of $\ln\Lambda_\tau(\mathbf{Z}_{1:\tau})$ over boundaries. Furthermore, equations (\ref{perm_condition}) are fulfilled so that the bounds on $\lambda_0$ and $\lambda_1$ of proposition \ref{prop_bounds} hold and the extrema of the ASN are (asymptotically) reached for $a(x,x)=1$ $\forall\; x\in\mathcal{S}$ and $a(x,y)=1/M$ $\forall\; x,y\in\mathcal{S}$ ($\sigma=10^{-1}$ and $\sigma=10^4$ given the adopted quantization step). It is confirmed then the intuitive idea that more compact priors allows easier detections.
\begin{figure}[t]
\centering
\includegraphics[width=\columnwidth]{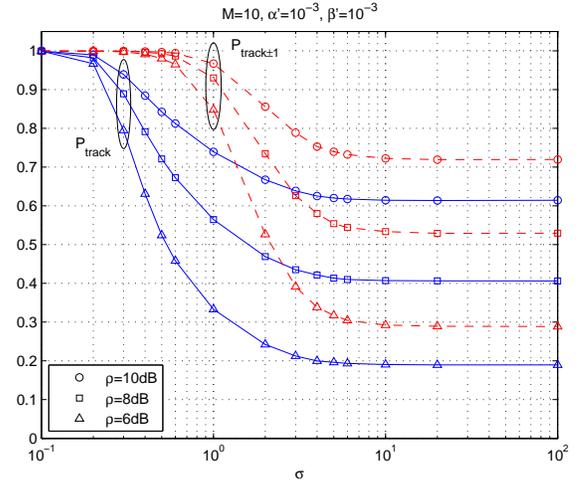}
\vspace{-0.7cm}\caption{Probabilities of correct track estimation $P_\text{track}$ and $P_{\text{track}\pm1}$ versus the mobility parameter $\sigma$ for different values of $\rho$ and for $M=11$.}\label{fig2}
\end{figure}
Figure \ref{fig2} shows the effect of the SNR on the probability of correct track estimation. It has been also plotted $P_{\text{track}\pm1}$, the probability that the distance between each state of the recovered trajectory and the actual state is less then or equal to 1. Notice that, since an uncoupled design has been adopted, the estimation performances decrease as $\rho$ is lowered and/or $\sigma$ is increased while $\alpha$  and $\beta$ are not influenced by these parameters (indeed lower values of $\rho$ and/or larger values of $\sigma$ are traded for larger ASNs).

The remaining curves concern more specifically the radar scenario outlined in section \ref{radar_appl_sec}. The search zone is composed of a single elevation and 4 azimuth sectors; the other parameters are $N_r=100$ and $N_d=16$. The transition probabilities along the third dimension (i.e. range) are defined as above, with a maximum admissible range transition of $\pm3$ bins. Azimuth transitions are neglected while Doppler ones are assumed equally likely (i.e. $\sigma_2=0$ and $\sigma_4=+\infty$). The truncation stage is $K=20$, with $\gamma_K=\sqrt{\gamma_0\gamma_1}$, while the nominal SNR has been set at $\rho'=N_d$\footnote{If each frame results from processing pulse trains of $N_d$ pulses, $\rho'=N_d$ corresponds to an SNR per pulse of 0 dB.} (notice that, since it is not realistic to assume prior knowledge of the target strength, the actual SNR is in general different from the design value $\rho'$). The subsequent plots are aimed both at assessing the effect of a mismatch between $\rho$ and $\rho'$ and at giving a comparison with an equivalent FSS procedure exhibiting the same $P_\text{fa}$.
\begin{figure}[t]
\centering
\includegraphics[width=\columnwidth]{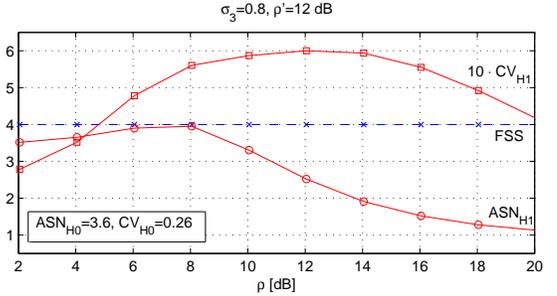}
\vspace{-0.7cm}\caption{ASN and coefficient of variation versus the SNR under $H_1$. The dotted line represent the sample size of the equivalent FSS rule.}\label{fig3}
\end{figure}
In figure \ref{fig3}, the ASN and the coefficient of variation (CV) of the sample size\footnote{The coefficient of variation of a random variable is the ratio $\sigma/\mu$ of its standard deviation $\sigma$ and its mean $\mu\neq0$.} are represented versus $\rho$ under $H_1$. Notice the characteristic peak at intermediate values of the SNR: yet the effect of the beam antenna remaining blocked monitoring a particular direction has been avoided by truncation. As for the FSS rule, a conservative choice has been made taking 4 samples, which is uniformly larger than the ASN of the truncated sequential procedure. Figure \ref{fig4} shows $P_\text{d}$ and $P_\text{track}$ versus the SNR for both the sequential and the FSS procedure. It can be seen that the sequential procedure achieve larger $P_\text{d}$ over all the inspected range of SNRs maintaining, at the same time, a full sample size saving. Notice also the massive gain granted in terms of $P_\text{track}$ which is mainly due to the low ASN required.
\begin{figure}[t]
\centering
\includegraphics[width=\columnwidth]{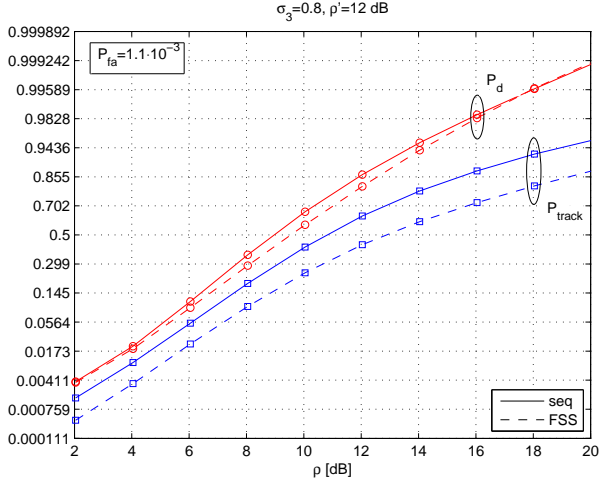}
\vspace{-0.7cm}\caption{Probability of detection and correct trajectory estimation versus the SNR for both the sequential and the FSS procedures. Vertical axis in Gaussian scale.}\label{fig4}
\end{figure}
Finally, figure \ref{fig5} shows the performances in terms of the target mobility. $P_\text{d}$, $P_\text{track}$, $P_{\text{track}\pm1}$ and the ASN are represented versus $\sigma_3$ for $\rho=\rho'$ (recall that $\sigma=10^{-1}$ corresponds to the case of a steady target). It can be seen, that, while the probability of detection of the FSS procedure impairs as the target mobility increases, that of the sequential rule remains almost unchanged, in that large values of $\sigma_3$ are counterbalanced by higher ASNs. As to the estimation performance, it obviously decreases in both cases, but sequential techniques retain their superiority over all the range of $\sigma_3$.

\section{Conclusions}\label{conclusions}
The general problem of sequential detection and possible trajectory estimation of a dynamic system observed through a set of noisy measurements has been considered. Previous limitation on the system dynamics imposed by other works present in the literature have been removed and a thorough analysis of the asymptotic behaviour of the test has been presented. In particular, it has been shown that, under rather mild conditions, the test is asymptotically optimal, in the sense that it minimizes, up to a $\mathscr{O}(1)$ term, any moment of the stopping time distribution under both hypothesis as the probabilities of error approach zero. Possible applications to radar surveillance problems have been inspected. Finally, the numerical has confirmed the correctness of the given approximations and has demonstrated the merits of the proposed strategy with respect to other competitors in the context of radar surveillance.
\begin{figure}[t]
\centering
\includegraphics[width=\columnwidth]{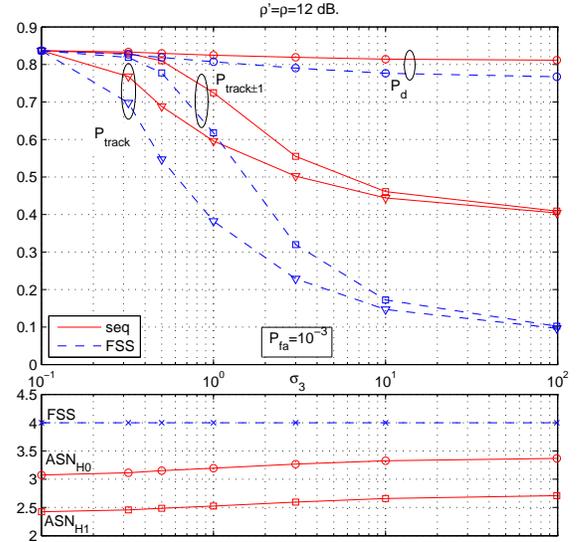}
\vspace{-1.5cm}\caption{Probabilities of detection and trajectory estimation versus the range target mobility parameter for both sequential and FSS procedures (top); ASN in the lower plot.}\label{fig5}
\end{figure}

\renewcommand{\thetheorem}{A.\arabic{theorem}}
\setcounter{equation}{0}
\renewcommand{\theequation}{A.\arabic{equation}}

\appendix

In the following, the proofs of theorems \ref{theorem_conv_as}, \ref{asym_optimality} and propositions \ref{prop_bound_sup}, \ref{prop_bounds} are given.

\begin{IEEEproof}[Proof of theorem \ref{theorem_conv_as}]
Part of the proof borrows its arguments from \cite{Ler1992}. Since condition \ref{asym_cond_1} is satisfied, $\{X_i\}_{i\in\mathbb{N}}$ is stationary and ergodic and, from \cite[lemma 1]{Ler1992}, $\{Z_i\}_{i\in\mathbb{N}}$ is stationary and ergodic as well. Given the one-sided stationary process $\big\{(X_i,Z_i)\big\}_{i\in\mathbb{N}}$, it is extended to a two-sided stationary process $\big\{(X_i, Z_i)\big\}_{i\in\mathbb{Z}}$ in the usual way.

Define
\begin{align}
q_k(\mathbf{z}_{1:k}|x)=&\frac{f(z_1|x)}{f(z_1|\theta_0)} \sum_{x_2\in\mathcal{S}}\cdots \sum_{x_k\in\mathcal{S}} a(x,x_2) \frac{f(z_2|x_2)}{f(z_2|\theta_0)}\cdot\notag\\
&\cdot \prod_{i=3}^k a(x_{i-1},x_i) \frac{f(z_i|x_i)}{f(z_i|\theta_0)},\notag\\
q_k(\mathbf{z}_{1:k})=&\max_{x\in\mathcal{S}}q_k(\mathbf{z}_{1:k}|x),\notag
\end{align}
with $q_1(z_1|x)=\frac{f(z_1|x)}{f(z_1|\theta_0)}$. Since $q_k(\mathbf{z}_{1:k}|x)$ is the likelihood ratio $\Lambda_k(\mathbf{z}_{1:k})$ given that
$X_1=x$, it results that, $\forall\,\mathbf{z}_{1:k}\in\mathcal{Z}^k$ and $k\in\mathbb{N}$, $\Lambda_k(\mathbf{z}_{1:k})\leq q_k(\mathbf{z}_{1:k})$ and
\begin{align}
&\Lambda_k(\mathbf{z}_{1:k})= \sum_{x\in\mathcal{S}}\pi(x)q_k(\mathbf{z}_{1:k}|x)\geq
\max_{x\in\mathcal{S}} \big\{\pi(x)q_k(\mathbf{z}_{1:k}|x) \big\} \geq \notag\\
&\geq\max_{x\in\mathcal{S}} \Big\{\min_{y\in\mathcal{S}}\big\{\pi(y)\big\}
q_k(\mathbf{z}_{1:k}|x) \Big\}=\min_{x\in\mathcal{S}} \big\{\pi(x)\big\}
q_k(\mathbf{z}_{1:k}).\notag
\end{align}
Combining the above inequalities on $\Lambda_k(\mathbf{z}_{1:k})$ it results that
\begin{equation}
\frac{1}{k}\ln \min_{x_1\in\mathcal{S}} \big\{\pi(x_1)\big\} \leq
\frac{1}{k}\ln \frac{\Lambda_k(\mathbf{z}_{1:k})}{q_k(\mathbf{z}_{1:k})}\leq 0,\notag
\end{equation}
$\forall\,\mathbf{z}_{1:k}\in\mathcal{Z}^k$ and $k\in\mathbb{N}$, which implies that
\begin{align}
&\lim_{k\rightarrow+\infty} \frac{1}{k} \ln \Lambda_k(\mathbf{Z}_{1:k})=
\lim_{k\rightarrow+\infty} \frac{1}{k} \ln q_k(\mathbf{Z}_{1:k}), \text{a.s. under } H_i, \notag\\
&\lim_{k\rightarrow+\infty} \frac{1}{k} \operatorname{E}_{H_i}\big[\ln
\Lambda_k(\mathbf{Z}_{1:k})\big]= \lim_{k\rightarrow+\infty} \frac{1}{k}
\operatorname{E}_{H_i}\big[\ln q_k(\mathbf{Z}_{1:k})\big],\notag
\end{align}
for $i=0,1$. As a consequence, it is sufficient to demonstrate the conclusions of the theorem
for $q_k(\mathbf{Z}_{1:k})$. The advantage of working with $q_k$ rather than $\Lambda_k$
is two-fold. First, $q_k$ does not depend upon the initial probability $\pi$
and, then, the second part of the theorem is demonstrated. The second advantage
descends from the following relationship
\begin{equation}
q_{s+t}(\mathbf{z}_{1:s+t})\leq q_{s}(\mathbf{z}_{1:s}) q_{t}(\mathbf{z}_{s+1:s+t}),\quad \forall \,
s,t\geq1,\label{lemma3_Ler}
\end{equation}
for any sequence $\{z_i\}_{i\in\mathbb{N}}$ (the proof is identical to that of \cite[lemma 3]{Ler1992}). Define now a doubly indexed sequence of random
variables $\{W_{st}\}_{t> s\geq0}$ by $W_{st}=\ln q_{t-s}(\mathbf{Z}_{s+1:t})$. With this definition, the stochastic process $\{W_{st}\}_{t> s\geq0}$ satisfies the following three properties.
\begin{enumerate}
\item From equation (\ref{lemma3_Ler}), $W_{st}\leq W_{su}+W_{ut}$, $\forall \;s<u<t$, i.e. it is a subadditive process.
\item By the stationarity of $\{Z_k\}_{k\in\mathbb{Z}}$, $\{W_{st}\}_{t>s\geq0}$ is stationary relative to the shift transformation $W_{st}\rightarrow W_{s+1\;t+1}$, i.e. $W_{st}$ and $W_{s+1\;t+1}$ have the same distribution.
\item By condition \ref{asym_cond_3}, for $i=0,1$, it results
\begin{align}
\operatorname{E}_{H_i}[W_{01}^+]=&\operatorname{E}_{H_i}\left[ \max_{x\in\mathcal{S}}\bigg( \ln \frac{f(Z_1|x)}{f(Z_1|\theta_0)}  \bigg)^+ \right]\leq \notag\\
\leq&\sum_{x\in\mathcal{S}}\operatorname{E}_{H_i}\left[ \bigg| \ln \frac{f(Z_1|x)}{f(Z_1|\theta_0)}  \bigg| \right]<+\infty.\notag
\end{align}
where $v^+=\max\{0,v\}$, $v\in\mathbb{R}$.
\end{enumerate}
By the Kingman's subadditive ergodic theorem \cite[theorems 1.5 and 1.8]{Kin1976}, a stochastic process satisfying these properties also satisfies the conclusions of the ergodic theorem, i.e.
\begin{enumrm}
\item $\lim_{k\rightarrow+\infty}k^{-1}W_{0k}=W<+\infty
$ exists almost surely;
\item
$\operatorname{E}[W]=\lim_{k\rightarrow+\infty}k^{-1}\operatorname{E}[W_{
0k}]$;
\item $W$ is degenerate if the process is ergodic.
\end{enumrm}
Thus, an application to $W_{0k}=\ln q_k(\mathbf{Z}_{1:k})$ gives (ergodicity carries over from
$\{Z_k\}_{k\in\mathbb{N}}$)
\begin{equation}
\lim_{k\rightarrow+\infty}\frac{1}{k}\operatorname{E}_{H_i}\big[\ln
q_k(\mathbf{Z}_{1:k})\big]=\lambda_i<+\infty, \quad i=0,1,\notag
\end{equation}
exists and
\begin{equation}
\lim_{k\rightarrow+\infty}\frac{1}{k}\ln q_k(\mathbf{Z}_{1:k})=\lambda_i, \quad \text{
a.s. under }H_i, \;i=0,1.\notag
\end{equation}
Since $k^{-1}\ln \Lambda_k(\mathbf{Z}_{1:k})$ and $k^{-1}\ln q_k(\mathbf{Z}_{1:k})$ have the same limiting behaviour, the proof of the theorem is complete if one demonstrates that
$\lambda_0$ is finite, strictly negative and that $\lambda_1$ is strictly
positive.

In order to prove that $\lambda_0$ is bounded also from below, first notice that $\Lambda_k(\mathbf{z}_{1:k})$ is a convex combination of the terms $f_k(\mathbf{z}_{1:k}|\mathbf{x}_{1:k})/f_k(\mathbf{z}_{1:k}|\theta_0)$, for any $\mathbf{z}_{1:k}\in\mathcal{Z}^k$ and $k\in\mathbb{N}$, and, thus,
\begin{align}
\ln \Lambda_k(\mathbf{z}_{1:k})\geq& \ln \min_{\mathbf{x}_{1:k}\in\mathcal{S}^k} \frac{f_k(\mathbf{z}_{1:k}|\mathbf{x}_{1:k})}{f_k\mathbf{z}_{1:k}|\theta_0)} =\notag\\
=& \sum_{n=1}^k \min_{x\in\mathcal{S}}\ln \frac{f(z_n|x)}{f(z_n|\theta_0)}= \sum_{n=1}^k \eta(z_n),\notag
\end{align}
where $\{\eta(Z_k)\}_{k\in\mathbb{N}}$ forms a sequence of i.i.d. random variables under $H_0$. Furthermore, by condition \ref{asym_cond_3}, it results that
\begin{equation}
\operatorname{E}_{H_0}\big[ |\eta(Z_1)| \big]=\operatorname{E}_{H_0}\Bigg[ \bigg|\min_{x\in\mathcal{S}}\ln \frac{f(Z_1|x)}{f(Z_1| \theta_0)} \bigg|\Bigg]<+\infty.\notag
\end{equation}
This implies that
\begin{equation}
\lambda_0=\lim_{k\rightarrow+\infty}\frac{1}{k}\operatorname{E}_{H_0}\big[\ln\Lambda_k(\mathbf{Z}_{1:k})\big]\geq \operatorname{E}_{H_0}\big[ \eta(Z_1) \big]>-\infty.\notag
\end{equation}

As for the sign of $\lambda_i$, $i=0,1$, let $g_n(z_n|\mathbf{z}_{1:n-1})$ denote the conditional density given by the ratio $f_{n,H_1}(\mathbf{z}_{1:n}) / f_{n-1,H_1}(\mathbf{z}_{1:n-1})$ for $n\geq2$ and by $f_{1,H_1}(z_1)$ for $n=1$. With this notation, the limiting constants $\lambda_i$ are also given by
\begin{align}
\lambda_i&=\lim_{k\rightarrow+\infty}\frac{1}{k}\operatorname{E}_{H_i}\big[
\ln\Lambda_k(\mathbf{Z}_{1:k})\big]=\notag\\
&=\lim_{k\rightarrow+\infty}\frac{1}{k}\sum_{n=1}^k\operatorname{E}_{H_i}\Bigg[
\ln\frac{g_n(Z_n|\mathbf{Z}_{1:n-1})}{f(Z_n|\theta_0)}\Bigg]=\notag\\
&=\lim_{k\rightarrow+\infty}\frac{1}{k}\sum_{n=1}^k\operatorname{E}_{H_i}\Bigg[
\ln\frac{g_n(Z_1|\mathbf{Z}_{-n+2:0})}{f(Z_1|\theta_0)}\Bigg],\quad i=0,1,\label{lambda_i}
\end{align}
where stationarity of $\{Z_i\}_{i\in\mathbb{Z}}$ has been exploited. On the other hand, $g_n(Z_1|\mathbf{Z}_{-n+2:0})$ can be written as $\sum_{x\in\mathcal{S}} f(Z_1|x) P\big(\{X_1=x\}|\mathbf{Z}_{-n+2:0}\big)$, $\forall \;n\in\mathbb{N}$, and this implies that
\begin{equation}
\min_{x\in\mathcal{S}} \frac{f(Z_1|x)}{f(Z_1|\theta_0)}\leq \frac{g_n(Z_1|\mathbf{Z}_{-n+2:0})}{f(Z_1|\theta_0)} \leq \max_{x\in\mathcal{S}}  \frac{f(Z_1|x)}{f(Z_1|\theta_0)},\;n\in\mathbb{N},\notag
\end{equation}
and, by condition \ref{asym_cond_3}, that $\left\{\ln \frac{g_n(Z_1| \mathbf{Z}_{-n+2:0})}{f(Z_1| \theta_0)}\right\}_{n\in\mathbb{N}}$ is a sequence of uniformly integrable random variables. In this case, dominated convergence gives
\begin{equation}
\lim_{n\rightarrow+\infty}  \operatorname{E}_{H_i}\left[\ln \frac{g_n(Z_1|\mathbf{Z}_{-n+2:0})} {f(Z_1|\theta_0)}\right] = \operatorname{E}_{H_i}\left[\ln \frac{g(Z_1|\mathbf{Z}_{-\infty:0})} {f(Z_1|\theta_0)}\right],\label{lim_LLR}
\end{equation}
$i=0,1$. In the above equation, $g(Z_1|\mathbf{Z}_{-\infty:0})$ denotes the limit
\begin{align}
&\lim_{n\rightarrow+\infty}g_n(Z_1|\mathbf{Z}_{-n+2:0})=\notag\\
&=\lim_{n\rightarrow+\infty}\sum_{x\in\mathcal{S}} f(Z_1|x) P\big(\{X_1=x\}|\mathbf{Z}_{-n+2:0}\big)=\notag\\
&=\sum_{x\in\mathcal{S}} f(Z_1|x) P\big(\{X_1=x\}|\mathbf{Z}_{-\infty:0}\big),\label{mixture_0}
\end{align}
where the latter equality follows form a martingale convergence theorem by L\'evy (see \cite{Shi1995}). Finally, from (\ref{lim_LLR}) and from the Ces\'aro mean theorem follow that $\lambda_i$ in equation (\ref{lambda_i}) can be also written as
\begin{equation}
\lambda_i= \operatorname{E}_{H_i}\left[\ln \frac{g(Z_1|\mathbf{Z}_{-\infty:0})} {f(Z_1|\theta_0)}\right], \quad i=0,1.\notag
\end{equation}
which implies that
\begin{subequations}\label{lambda_sign}
\begin{align}
\lambda_1=&\operatorname{E}_{H_1}\left[ \operatorname{D} \big(
g(\cdot|\mathbf{Z}_{-\infty:0})\Vert f(\cdot|\theta_0)\big)\right]>0,\\
\lambda_0=&-\operatorname{E}_{H_0}\left[ \operatorname{D} \big(f(\cdot|\theta_0) \Vert
g(\cdot|\mathbf{Z}_{-\infty:0}) \big)\right]<0.
\end{align}
\end{subequations}
Inequalities in (\ref{lambda_sign}) descend from the fact that the Kullback-Leibler divergence is always non negative and is equal to zero if and only if the two densities are equal $\nu$-a.e.: this, however, cannot happen since, from equation (\ref{mixture_0}), $g\big(\cdot|\{\mathbf{Z}_{-\infty:0} = \mathbf{z}_{-\infty:0}\}\big)$ is always a mixture of $M$ elements of $\{f(\cdot|x)\}_{x\in\mathcal{S}}$, which is not $\nu$-a.e. equal to $f(\cdot|\theta_0)$ by condition \ref{asym_cond_2}.
\end{IEEEproof}

\begin{IEEEproof}[Proof of theorem \ref{asym_optimality}]

Exploiting the idea introduced in \cite{Fuh2003, Fuh2004}, the likelihood ratio can be equivalently represented as $\Lambda_k(\mathbf{Z}_{1:k})=\left\Vert \matr{M}_k \vec{\pi} \right \Vert$, $k\in\mathbb{N}$, where $\Vert \cdot \Vert$ is the $L_1$-norm on $\mathbb{R}^M$ and $\matr{M}_k$ is a $M\times M$ matrix on $\mathbb{R}$ defined as follows
\begin{equation}
\matr{M}_1=\matr{T}_1,\quad \matr{M}_k=\matr{T}_k \matr{A}^T \matr{M}_{k-1}, \quad \text{for } k\geq 2,\notag
\end{equation}
$(\cdot)^T$ denoting transpose, $\matr{T}_k$ being a diagonal $M \times M$ matrix with entries $\big\{f(Z_k|x)/ f(Z_k|\theta_0)\}_{x\in\mathcal{S}}$, $\matr{A}=[a(x,y)]_{x,y\in\mathcal{S}}$ the transition probability matrix and $\vec{\pi}=[\pi(x)]_{x\in\mathcal{S}}$ the initial probability vector. Under conditions \ref{asym_cond_5} and \ref{asym_cond_6}, $\matr{M}_k$ is invertible for every $\mathbf{z}_{1:k}\in\mathcal{Z}^k$, $k\in\mathbb{N}$, while, under hypothesis $H_1$, $\{X_k,Z_k\}_{k\in\mathbb{N}}$ is a Markov chain on $\mathcal{S}\times \mathcal{Z}$. This implies that the process $\big\{(X_k,Z_k),\matr{M}_k\big\}_{k\in\mathbb{N}}$ is a multiplicative Markovian Process (see \cite[definition 1.1]{Bou1988}) and $\big\{\matr{M}_k \big\}_{k\in\mathbb{N}}$ a product of Markov random matrices.

In order to exploit the large deviations result for products of Markov random matrices in \cite{Bou1988}, it is first needed to verify the validity of conditions $A$ of \cite{Fuh2003} and \cite{Bou1988}.
Under condition \ref{asym_cond_1}, $\{X_k\}_{k\in\mathbb{N}}$ is uniformly ergodic and so is the Markov chain $\big\{(X_k,Z_k)\big\}_{k\in\mathbb{N}}$ (see \cite{Cap2005}), i.e. condition $A_1$ is fulfilled. As concerns $A_2$, it is requested that there exists $p>0$ such that
\begin{equation}
\operatorname{E}\left[e^{p\sup\left \{\ln \left\Vert \matr{M}_k\right \Vert, \ln \left\Vert \matr{M}_k^{-1} \right\Vert \right\}} \big| \, X_1=x_1, Z_1=z_1\right]<+\infty,\label{A2_cond}
\end{equation}
$\forall \, (x_1,z_1)\in\mathcal{S}\times\mathcal{Z}$ and $k=0,1$, where the expectation is taken with respect to the joint distribution of $\{X_k,Z_k\}_{k\in\mathbb{N}}$ and the matrix norm is that induced by the vector norm, i.e. $\big\Vert \matr{M}_k \big\Vert=\sup_{\Vert \vec{u} \Vert =1} \left\Vert\matr{M}_k\vec{u}\right\Vert$. Now it results that
\begin{align}
&e^{p\sup\left \{\ln \left\Vert \matr{M}_2\right\Vert, \ln \left\Vert  \matr{M}_2^{-1} \right \Vert \right \} }\leq \big\Vert \matr{T}_2\matr{A}^T \matr{T}_1 \big \Vert^p + \big\Vert \big( \matr{T}_2\matr{A}^T \matr{T}_1\big)^{-1} \big \Vert^p\!\leq\notag\\
&\leq \left(\big\Vert \matr{A}^T \big \Vert \big\Vert \matr{T}_1 \big \Vert \big\Vert \matr{T}_2\big \Vert\right)^p + \left(\big\Vert \big( \matr{A}^T \big)^{-1}\big \Vert \big\Vert \matr{T}_1^{-1}\big \Vert \big\Vert \matr{T}_2^{-1}\big \Vert\right)^p, \notag
\end{align}
with
\begin{equation}
\big\Vert \matr{T}_k^i\big \Vert^p= \left( \max_{x\in\mathcal{S}}\left( \frac{f(Z_k|x)}{f(Z_k|\theta_0)}\right)^i \right)^p \leq \sum_{x\in\mathcal{S}}\left( \frac{f(Z_k|x)}{f(Z_k|\theta_0)}\right)^{i p},\notag
\end{equation}
$i=\pm1$. This implies that equation (\ref{A2_cond}) is satisfied if
\begin{equation}
\operatorname{E}\left[ \left( \frac{f(Z_1|w)}{f(Z_1|\theta_0)}\frac{f(Z_2|x)}{f(Z_2|\theta_0)}\right)^{\pm p} | \, X_1=x_1, Z_1=z_1\right]<+\infty,\notag
\end{equation}
$\forall\, w,x,x_1\in\mathcal{S}$ and $z_1\in\mathcal{Z}$. But
\begin{align}
&\operatorname{E}\left[ \left( \frac{f(Z_1|w)}{f(Z_1|\theta_0)} \frac{f(Z_2|x)}{f(Z_2|\theta_0)}\right)^{\pm p} | \, X_1=x_1, Z_1=z_1\right]= \notag\\
&=\left( \frac{f(z_1|w)}{f(z_1|\theta_0)}\right)^{\pm p}\! \sum_{y\in\mathcal{S}} \int_\mathcal{Z}\!\!\left( \frac{f(z|x)}{f(z|\theta_0)}\right)^{\pm p} \!\!\!a(x_1,y) f(z|y) \nu(dz)=\notag\\
&=\left( \frac{f(z_1|w)}{f(z_1|\theta_0)}\right)^{\pm p} \sum_{y\in\mathcal{S}} a(x_1,y) \operatorname{E}_y\left[ \left( \frac{f(Z_1|x)}{f(Z_1 |\theta_0)}\right)^{\pm p} \right],\notag
\end{align}
which is bounded for every $(w,z_1)\in\mathcal{S} \times \mathcal{Z}$ by condition \ref{asym_cond_5} and for every $x_1,x,y\in\mathcal{S}$ by condition \ref{asym_cond_4}. As concerns condition $A_3$, the process $\big\{(X_k,Z_k),\matr{M}_k \big\}_{k \in\mathbb{N}}$ has to be strongly irreducible and contracting (see \cite[definition 2]{Fuh2003} for the terminology) and this can be shown using arguments similar to those in \cite[proof of proposition 4]{Fuh2003}.

Since these three conditions are satisfied, it results that:
\begin{enumrm}
\item\label{Boug1} there exists a neighbourhood $I$ of the origin such that, for every $p\in I$, $\Big(\operatorname{E}\Big[ \left\Vert \matr{M}_k \vec{\pi}\right\Vert^p | $ $\, X_1=x_1, Z_1=z_1\Big]\Big)^{1/k}$ converges to a function $H(p)$ uniformly in $(x_1,z_1)\in \mathcal{S}\times \mathcal{Z}$ \cite[section 4 and theorem 4.3]{Bou1988};
\item\label{Boug2} $H'(0)=\lambda_1$ \cite[proposition 3.8]{Bou1988}.
\end{enumrm}
These two properties will be used to derive the convergence rate of the sequence $\{k^{-1}\ln\Lambda_k(\mathbf{Z}_{1:k}) \}_{k\in\mathbb{N}}$. To this end, define the function
\begin{align}
\widetilde{H}(p)=& \limsup_{k\rightarrow+\infty}\frac{1}{k} \ln\operatorname{E}_{H_1}\left[e^{p \ln\Lambda_k (\mathbf{Z}_{1:k})}\right] =\notag\\
=& \limsup_{k\rightarrow+ \infty}\frac{1}{k}\ln \operatorname{E}_{H_1}\left[ \Lambda_k^p (\mathbf{Z}_{1:k})\right].\notag
\end{align}
Given the uniform convergence in $(x_1,z_1)$ and recalling that $\left\Vert \matr{M}_k \vec{\pi} \right\Vert = \Lambda_k(\mathbf{Z}_k)$, property (\ref{Boug1}) implies that
\begin{equation}
\lim_{k\rightarrow+ \infty}\frac{1}{k}\ln \operatorname{E}_{H_1}\left[ \Lambda_k^p (\mathbf{Z}_{1:k})\right]=\ln H(p),\; \forall\,p\in I,\notag
\end{equation}
and then $\widetilde{H}(p)= \ln H(p)$, $\forall\,p\in I$, so that, from property (\ref{Boug2}), $\widetilde{H}'(0)=\lambda_1$. Denote now with $\mathcal{D}_{\widetilde{H}}$ the set $\{ p\in\mathbb{R}:\; \widetilde{H}(p)<+\infty \}$. Since, from condition \ref{asym_cond_4},
\begin{align}
\operatorname{E}_{H_1}\left[ \Lambda_k^a (\mathbf{Z}_{1:k})\right]\leq& \operatorname{E}_{H_1}\left[ \prod_{n=1}^k\max_{x\in\mathcal{S}}\left(\frac{f(Z_n| x)}{f(Z_n|\theta_0)}\right)^a \right]\leq\notag\\
\leq& \left(\sum_{y\in\mathcal{S}}\sum_{x\in\mathcal{S}}\operatorname{E}_y\left[ \left(\frac{f(Z_1| x)}{f(Z_1|\theta_0)}\right)^a\right]  \right)^k<+\infty,\notag
\end{align}
it follows that $\widetilde{H}(a)<+\infty$ and, thus, the interior part of $\mathcal{D}_{\widetilde{H}}$ contains the point $p=0$. This and the fact that $\widetilde{H}'(0)=\lambda_1$ implies that $\big\{\frac{1}{k} \ln \Lambda_k(\mathbf{Z}_{1:k}) \big\}_{k\in\mathbb{N}}$ converges to $\lambda_1$ exponentially in $k$ \cite[exercise 2.3.25]{Dem1998}, \cite[theorem IV.1]{Ell1984}.\footnote{A sequence $\{Y_k\}_{k\in\mathbb{N}}$ of random variables is said to converge exponentially to a constant $\lambda$ if, for any sufficiently small $\epsilon>0$, there exists a constant C such that $P\big(|Y_k-\lambda|\geq\epsilon\big)\leq e^{-kC}$\cite{Ell1984}.}

The exponential convergence is obviously much stronger than the a.s. convergence granted by theorem \ref{theorem_conv_as}. Indeed, the former implies that
\begin{equation}
\sum_{k=1}^{+\infty}k^r P\left(\Big\{ \big|k^{-1} \ln \Lambda_k(\mathbf{Z}_{1:k})  - \lambda_1\big|\geq \varepsilon \Big\}|H_1\right)< +\infty, \notag
\end{equation}
$\forall\, \varepsilon>0$ and $r>0$, which in turn implies that $\big\{k^{-1} \ln \Lambda_k(\mathbf{Z}_{1:k}) \big\}_{k\in\mathbb{N}}$ converges $r$-quickly to $\lambda_1$ for any $r>0$ \cite[lemma 3]{Lai1975}.\footnote{A sequence $\{Y_k\}_{k\in\mathbb{N}}$ of random variables is said to converge $r$-quickly to a constant $\lambda$, for some $r>0$, if $\operatorname{E}[T^r_\epsilon]<+\infty$, for all $\epsilon>0$, where $T_\epsilon=\sup\{k\in\mathbb{N}:\; |Y_k-\lambda|\geq\epsilon\}$ ($\sup\{\emptyset\}=0$) \cite{Lai1977}.} The latter is also called strong complete convergence in \cite{Tar1999,Dra1999}. Given the $r$-quick convergence for any $r>0$, since $\lambda_1$ is finite and strictly positive from theorem \ref{theorem_conv_as}, the thesis follows from \cite[corollary 1]{Lai1981}.

The case under $H_0$ can be handled similarly considering $\left\{Z_k, \matr{M}_k\right\}_{k\in\mathbb{N}}$, where $\{Z_k\}_{k\in\mathbb{N}}$ is now an i.i.d. process.
\end{IEEEproof}

In order to prove propositions \ref{prop_bound_sup} and \ref{prop_bounds} the following lemma is first needed.
\begin{lemma}\label{convexity_div}
$ \operatorname{D} \left(\sum_{x\in\mathcal{S}}c(x)f(\cdot|x) \;\Vert\; f(\cdot|\theta_0) \right)$ and $\operatorname{D} \left( f(\cdot|\theta_0)\;\Vert\; \sum_{x\in\mathcal{S}}c(x)f(\cdot|x)\right)$ are convex function on the set $\big\{\{c(x)\}_{x\in\mathcal{S}}\in[0,1]^M:\;\sum_{x\in\mathcal{S}}c(x)=1\big\}$.
\end{lemma}
\begin{IEEEproof}
It can be verified exploiting Jensen inequality and convexity of functions $-\ln v$ and $v\ln v$, $v\in\mathbb{R}^+$.
\end{IEEEproof}

\begin{IEEEproof}[Proof of proposition \ref{prop_bound_sup}]
From the proof of theorem \ref{theorem_conv_as}, equations (\ref{mixture_0}) and (\ref{lambda_sign}), $\lambda_1$ can be also written as $\operatorname{E}_{H_1}\Big[ \operatorname{D} \big(
\sum_{x\in\mathcal{S}} f(\cdot|x) P\big(\{X_1=x\}|\mathbf{Z}_{-\infty:0}\big)\Vert f(\cdot|\theta_0)\big)\Big]$. On the other hand, lemma \ref{convexity_div} and Jensen inequality allow to write
\begin{align}
\textstyle \operatorname{D} \Big( \sum\limits_{x\in\mathcal{S}} f(\cdot|x) P\big(\{X_1=x\}|\{\mathbf{Z}_{-\infty:0} =\mathbf{z}_{-\infty:0}\}\big)\Vert f(\cdot|\theta_0)\Big)\leq \notag\\
\leq\sum_{x\in\mathcal{S}} P\big(\{X_1=x\}|\{\mathbf{Z}_{-\infty:0} =\mathbf{z}_{-\infty:0}\}\big)\operatorname{D} \big(  f(\cdot|x)\Vert f(\cdot|\theta_0)\big),\notag
\end{align}
for every realization $\mathbf{z}_{-\infty:0}$ of $\mathbf{Z}_{-\infty:0}$, whereby
\begin{align}
\lambda_1\leq&\operatorname{E}_{H_1}\left[ \sum_{x\in\mathcal{S}} P\big(\{X_1=x\}|\mathbf{Z}_{-\infty:0}\big)\operatorname{D} \big(  f(\cdot|x)\Vert f(\cdot|\theta_0)\big)\right]=\notag\\
=& \sum_{x\in\mathcal{S}} \operatorname{E}_{H_1}\Big[P\big(\{X_1=x\}|\mathbf{Z}_{-\infty:0}\big)\Big]\operatorname{D} \big(  f(\cdot|x)\Vert f(\cdot|\theta_0)\big)=\notag\\
=& \sum_{x\in\mathcal{S}}\overline{\pi}(x)D\big(f(\cdot|x)\Vert f(\cdot|\theta_0) \big).\notag
\end{align}
The upper bound on $|\lambda_0|$ can be proved similarly.
\end{IEEEproof}

\begin{IEEEproof}[Proof of proposition \ref{prop_bounds}]
From (\ref{perm_condition}), setting $c(x)=1$ for some $x\in\mathcal{S}$, it follows that $\operatorname{D} \big(  f(\cdot|x)\Vert f(\cdot|\theta_0)\big)$ has the same value for every $x\in\mathcal{S}$: this, along with proposition \ref{prop_bound_sup}, demonstrates the upper bound on $\lambda_1$. As to the lower bound, exploiting lemma \ref{convexity_div} and Jensen inequality, it follows that
\begin{align}
&\operatorname{D} \left( {\textstyle\sum_{x\in\mathcal{S}}} c(x) f(\cdot|x)\Vert f(\cdot|\theta_0)\right)=\notag\\
&=\frac{1}{M!} \sum_{i=1}^{M!} \operatorname{D} \left( {\textstyle\sum_{x\in\mathcal{S}}} c'_i(x) f(\cdot|x)\Vert f(\cdot|\theta_0)\right)\geq\notag\\
&\textstyle \geq \operatorname{D} \left( \sum_{x\in\mathcal{S}}  f(\cdot|x) \frac{1}{M!}\sum_{i=1}^{M!} c'_i(x)\Vert f(\cdot|\theta_0)\right)=\notag\\
&=\operatorname{D} \left( \sum_{x\in\mathcal{S}} \frac{1}{M}f(\cdot|x)\Vert f(\cdot|\theta_0)\right),\label{min_divergence}
\end{align}
for every probability vector $c$ on $\mathcal{S}$, where $\{c'_i\}_{i=2}^{M!}$, is the set of all of the possible permutations of $c$ and $c'_1=c$. From the demonstration of theorem \ref{theorem_conv_as}, equations (\ref{mixture_0}) and (\ref{lambda_sign}), $\lambda_1$ can be written also as $\operatorname{E}_{H_1}\Big[ \operatorname{D} \left( \sum_{x\in\mathcal{S}} f(\cdot|x) P\big(\{X_1=x\}|\mathbf{Z}_{-\infty:0}\big) \Vert f(\cdot|\theta_0)\right)\Big]$, and thus, exploiting (\ref{min_divergence}), it results that
\begin{align}
\lambda_1 \geq& \textstyle\operatorname{E}_{H_1}\Big[\operatorname{D} \left( \sum_{x\in\mathcal{S}} \frac{1}{M}f(\cdot|x)\Vert f(\cdot|\theta_0)\right)\Big]=\notag\\
=&\textstyle \operatorname{D} \left( \sum_{x\in\mathcal{S}} \frac{1}{M}f(\cdot|x)\Vert f(\cdot|\theta_0)\right),\notag
\end{align}
and the lower bound is proved. The bounds on $|\lambda_0|$ can be proved similarly.
\end{IEEEproof}

\bibliographystyle{IEEEtran}
\bstctlcite{bibliography:BSTcontrol}

\begin{IEEEbiography}{Emanuele Grossi}
was born in Sora, Italy on May 10, 1978. He received with honors the Dr. Eng. degree in Telecommunication Engineering in 2002 and the Ph.D. degree in Electrical Engineering in 2006, both from the University of Cassino, Italy. From February 2005 he spent six month at the Department of Electrical \& Computer Engineering of the University of British Columbia, Vancouver, as a Visiting Scholar. Since February 2006, he is assistant professor at the University of Cassino. His research interests concern wireless multiuser communication systems, radar detection and tracking, and statistical decision problems with emphasis on sequential analysis.
\end{IEEEbiography}

\begin{IEEEbiography}{Marco Lops}
was born in Naples, Italy on March 16, 1961. He received the Dr. Eng. degree in Electronic Engineering from the University of Naples in 1986.\\
From 1986 to 1987 he was in Selenia, Roma, Italy as an engineer in the Air Traffic Control Systems Group. In 1987 he joined the Department of Electronic and Telecommunications Engineering of the University of Naples as a Ph.D. student in Electronic Engineering. He received the Ph.D. degree in Electronic Engineering from the University of Naples in 1992. \\
From 1991 to 2000 he has been an Associate Professor of Radar Theory and Digital Transmission Theory at the University of Naples, while, since March 2000, he has been a Full Professor at the University of Cassino, engaged in research in the field of statistical signal processing, with emphasis on Radar Processing and Spread Spectrum Multiuser Communications. He also held teaching positions at the University of Lecce and, during $1991$, $1998$ and $2000$, he was on sabbatical leaves at University of Connecticut, Rice University, and Princeton University, respectively.
\end{IEEEbiography}

\end{document}